\documentclass[twocolumn,aps,prl,showpacs]{revtex4}

\usepackage{graphicx}% Include figure files
\newcommand{\id}{{\sf 1 \hspace{-0.3ex}\rule{0.1ex}{1.52ex}\rule[-.02ex]{0.3ex}{0.1ex} }}

\begin{document}
\title{Proposed Experiment to Test the Bounds of Quantum Correlations}
\author{Ad\'{a}n Cabello}
\email{adan@us.es}
\affiliation{Departamento de F\'{\i}sica Aplicada II,
Universidad de Sevilla, 41012 Sevilla, Spain}
\date{\today}
%First version: July 30, 2003.
%This version: February 6, 2004.
%After PRL proofs.

%%%%%%%%%%%%%%%%%%%%%%%%%%%% Abstract %%%%%%%%%%%%%%%%%%%%%%%%%%%%

\begin{abstract}
The combination of quantum correlations appearing in the
Clauser-Horne-Shimony-Holt inequality can give values between the
classical bound, $2$, and Tsirelson's bound, $2 \sqrt 2$. However,
for a given set of local observables, there are values in this
range which no quantum state can attain. We provide the analytical
expression for the corresponding bound for a parametrization of
the local observables introduced by Filipp and Svozil, and
describe how to experimentally trace it using a source of singlet
states. Such an experiment will be useful to identify the origin
of the experimental errors in Bell's inequality-type experiments
and could be modified to detect hypothetical correlations beyond
those predicted by quantum mechanics.
\end{abstract}

%%%%%%%%%%%%%%%%%%%%%%%%%%%%%%%%%%%%%%%%%%%%%%%%%%%%%%%%%%%%%%%%%%%

%Foundations of quantum mechanics; measurement theory
\maketitle

%%%%%%%%%%%%%%%%%%%%%%%%%%%%% Introduction %%%%%%%%%%%%%%%%%%%%%%%%%%%%%

Quantum mechanics is the most accurate and complete description of
the world known. This belief is supported by thousands of
experiments. Particularly, it is widely agreed that, leaving aside
some loopholes~\cite{Aspect99,Grangier01}, no local-realistic
theory of the type suggested by Einstein, Podolsky, and
Rosen~\cite{EPR35} is compatible with the experimental results
showing violations of Bell's inequalities~\cite{Bell64} and ``good
agreement'' with the predictions of quantum
mechanics~\cite{ADR82,SA88,OM88,OPKP92,TRO94,KMWZSS95,TBZG98,WJSHZ98,RKVSIMW01}.

On the other hand, current technology allows us to perform Bell's
inequality-type experiments with relative ease. Here we propose
using this possibility for a systematic test of the bounds of
quantum correlations. The benefits of this proposal, apart from
confirming quantum mechanics, would be to help us to identify and
discriminate between two sources of possible errors in Bell's
inequality-type experiments and to describe a method for searching
for hypothetical correlations beyond those predicted by quantum
mechanics.

To introduce the bounds of quantum correlations, let us consider a
high number of copies of systems of two distant particles prepared
in an unspecified way. Let $A$ and $a$ ($B$ and $b$) be physical
observables taking values $-1$ or $1$ referring to local
experiments on particle $I$ ($II$). Here we shall initially assume
that the particles have spin-$\frac{1}{2}$, and that $A$ is a spin
measurement along the direction represented by the unit ray $\vec
A$, etc. Fine~\cite{Fine82a} proved (see
also~\cite{GM82,Fine82b,WW01}) that a set of four correlation
functions $X_0 = \langle AB \rangle$, $X_1 = \langle Ab \rangle$,
$X_2 = \langle aB \rangle$, $X_3 = \langle ab \rangle$ can be
attained by a local-realistic theory (i.e., a theory in which the
local variables of a particle determine the results of local
experiments on this particle) if and only if they satisfy the
following eight Clauser-Horne-Shimony-Holt (CHSH)
inequalities~\cite{CHSH69}:
\begin{equation}
-2 \le X_i + X_{(i+1){\rm mod}\,4} + X_{(i+2){\rm mod}\,4} -
X_{(i+3){\rm mod}\,4} \le 2,
\label{CHSH}
\end{equation}
where $i=0,1,2,3$ and $(p+q){\rm mod}\,4$ means addition of $p$
and $q$ modulo $4$.

On the other hand, Tsirelson~\cite{Tsirelson80,Landau87} showed
that, for {\em any} quantum state $\rho$, the corresponding
quantum correlations, $x_0 = \langle AB \rangle_\rho$, $x_1 =
\langle Ab \rangle_\rho$, $x_2 = \langle aB \rangle_\rho$, $x_3 =
\langle ab \rangle_\rho$ must satisfy
\begin{eqnarray}
-1 & \le & x_i \le 1, \nonumber \\
-2 \sqrt{2} & \le & x_i + x_{(i+1){\rm mod}\,4} + x_{(i+2){\rm
mod}\,4} \nonumber \\
& & - x_{(i+3){\rm mod}\,4} \le 2 \sqrt{2}. \label{Tsirelson}
\end{eqnarray}
Quantum mechanics predicts violations of the CHSH inequalities
(\ref{CHSH}) up to $2 \sqrt{2}$. Such violations can be obtained
with pure~\cite{CHSH69} or mixed states~\cite{BMR92}.

However, inequalities~(\ref{Tsirelson}) are only a necessary but
not sufficient condition for the correlations to be attainable by
quantum mechanics. To illustrate this point, let us consider a set
of four numbers $Y_i$, such that they satisfy~(\ref{Tsirelson})
but not~(\ref{CHSH}); that is, their CHSH combinations lie between
Bell's classical bound and Tsirelson's bound. The question is: Are
there always a quantum state $\rho$ and four local observables
$A$, $a$, $B$, and $b$ such that $Y_i=x_i$? The answer is no;
certain sets of correlations {\em cannot} be reached by any
quantum state and any set of local observables. If quantum
mechanics is correct, this means that certain sets of expectation
values will never be found experimentally. Therefore, the notion
of superquantum correlations (i.e., correlations beyond those
predicted by quantum mechanics) is not only restricted to sets of
correlations such that the value of their CHSH operator is between
$2 \sqrt{2}$ and $4$ (the maximum possible value for the CHSH
operator)~\cite{PR94}, but it also covers some sets of
correlations whose CHSH operator is between $2$ and $2 \sqrt{2}$.

Uffink's quadratic inequalities~\cite{Uffink02} provide a more
restrictive necessary (but still not sufficient) condition for the
correlations to be attainable by quantum mechanics. The necessary
and sufficient condition for a set of four numbers to be reached
by quantum mechanics was found by Landau~\cite{Landau88} and
Tsirelson~\cite{Tsirelson93}, and has been rediscovered by
Masanes~\cite{Masanes03}. Four numbers $y_i$ can be reached by a
quantum state and some local observables (i.e., $y_i=x_i$) if and
only if they satisfy a the following eight inequalities:
\begin{eqnarray}
\!\!\!\!-\pi\! & \le &\! \arcsin{y_i} + \arcsin{y_{(i+1){\rm mod}\,4}} \nonumber \\
\!\!\!\!& &\! + \arcsin{y_{(i+2){\rm mod}\,4}} -
\arcsin{y_{(i+3){\rm mod}\,4}} \le \pi. \label{Masanes}
\end{eqnarray}
The inequalities (\ref{Masanes}) define the whole set of quantum
correlations but do not provide a practical characterization of
its bounds.

A different approach has been proposed by Filipp and
Svozil~\cite{FS03}. They define the quantum bounds as follows: Let
us choose several particular sets of local observables
$\{A_j,a_j,B_j,b_j\}$; let us use a computer to randomly generate
a high number of arbitrary quantum states $\{\rho_k\}$, and
calculate for all of them the value of the CHSH operator defined
as
\begin{equation}
{\rm CHSH} = \langle AB \rangle_\rho + \langle Ab \rangle_\rho +
\langle aB \rangle_\rho - \langle ab \rangle_\rho.
\label{CHSHoperator}
\end{equation}
The maximal and minimal values obtained are a numerical estimation
of the bounds of the quantum correlations. Given the way in which
the bounds have been constructed, for a given set of local
observables no quantum state (and, presumably, no other
preparation of physical systems) gives values outside these
bounds. A suitable parametrization of both the set of local
observables and the set of initial states yields to an analytical
expression of the bound of quantum correlations. However, Filipp
and Svozil's results are limited to a computer exploration of
these bounds. They conclude by saying that ``the exact analytical
geometries of quantum bounds remain unknown''~\cite{FS03}. In this
Letter, we provide the analytical expression of the bounds of the
quantum correlations using Filipp and Svozil's parametrization for
the local observables. We then use this analytical expression to
describe how to experimentally trace this bound. This experimental
verification will require a set of Bell's inequality-type tests,
each of them using a particular set of local observables and a
particular initial state. Both the local observables and the
initial state will depend on a single parameter $\theta$.

A suitable parametrization of the local observables and the
initial states should reflect the essential features of the bound
of quantum correlations: For every possible value of the
parameters, the CHSH operator should give values in the range
$[2,2 \sqrt{2}]$ (that is, between the classical Bell's bound and
Tsirelson's bound), but it will cover only a subset of the whole.
The area of this subset divided by the area of the whole set
should reflect the ratio between the difference between the
hypervolume of the convex set of quantum correlations defined by
(\ref{Masanes}) and that of the classical correlation polytope (a
four-dimensional octahedron)~\cite{Pitowsky86,Pitowsky89} defined
by (\ref{CHSH}) divided by the difference between the hypervolume
of the set defined by (\ref{Tsirelson}) (which is the intersection
of a bigger four-dimensional octahedron and a four-dimensional
cube) and that of the classical correlation polytope. Another
important property is that quantum bounds can always be attained
using a suitably chosen maximally entangled state. For practical
reasons, it would be interesting that the parametrization use as
few parameters as possible.

Filipp and Svozil~\cite{FS03} choose the following set of local
observables which depends on a single parameter,
\begin{eqnarray}
A & = & \cos (2 \theta) \sigma_z + \sin (2 \theta) \sigma_x,
\label{A} \\
B & = & \cos (\theta) \sigma_z + \sin (\theta) \sigma_x,
\label{B} \\
a & = & \sigma_z,
\label{a} \\
b & = & \cos (3 \theta) \sigma_z + \sin (3 \theta) \sigma_x,
\label{b}
\end{eqnarray}
where $0 \le \theta \le \pi$, and $\sigma_z$ and $\sigma_x$ are
the usual Pauli matrices.

%%%%%%%%%%%%%%%%%%%%%%%%%%%% Figure 1 %%%%%%%%%%%%%%%%%%%%%%%%%%%%%

\begin{figure}
\centerline{\includegraphics[width=8.2cm]{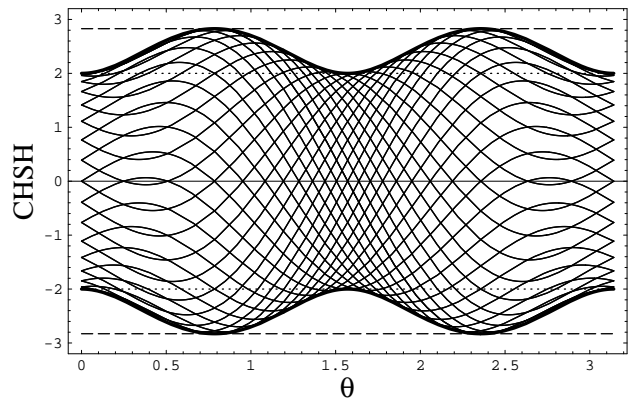}}
\caption{\label{FS02} Values of the CHSH operator for the
maximally entangled states $|\phi (\xi) \rangle = \cos\xi
|\phi^+\rangle+\sin\xi |\psi^-\rangle$, for $0 \le \xi < 2 \pi$,
for the local measurements (\ref{A})--(\ref{b}) parametrized by
$\theta$. The analytical expression of the quantum bound (thick
line) is given by (\ref{quantumbound}). Given a particular set of
local measurements (i.e., a particular value of $\theta$), the
quantum bound is reached by the maximally entangled states with
$\xi(\theta)$ given by (\ref{xi}). No other quantum state (and,
presumably, no other preparation of physical systems) gives values
outside these bounds. Dashed lines represent Bell's classical
bounds ($\pm 2$) and Tsirelson's bounds ($\pm 2 \sqrt{2}$).}
\end{figure}

%%%%%%%%%%%%%%%%%%%%%%%%%%%%%%%%%%%%%%%%%%%%%%%%%%%%%%%%%%%%%%%%%%%%

To obtain the analytical expression of the corresponding quantum
bound $F(\theta)$, it is useful to remember that, for each
$\theta$, the bound of quantum correlations can be reached by a
maximally entangled state. For the Filipp and Svozil
parametrization, a suitable set of maximally entangled states
turns out to be
\begin{equation}
|\varphi (\xi) \rangle = \cos\xi \, |\phi^+\rangle+\sin\xi \,
|\psi^-\rangle,
\label{frontierstate}
\end{equation}
where $0 \le \xi < 2 \pi$. In Fig.~\ref{FS02} we show the values
of the CHSH operator for several of these states. If, for a given
$\theta$, we calculate the maximum and minimum values of the CHSH
operator for the states $|\varphi (\xi) \rangle$, we obtain the
analytical expression for the bound numerically estimated
in~\cite{FS03}. The analytical expression of the bound of quantum
correlations is
\begin{eqnarray}
F(\theta) & = & \pm 2 \left\{ \left[1+\sin^2(2
\theta)\right]^{-1/2} \right. \nonumber \\
& & \left. +g(\theta) \sin(2 \theta) \left[1+{2 \over \cos(4
\theta)-3}\right]^{1/2}\right\},
\label{quantumbound}
\end{eqnarray}
where
\begin{equation}
g(\theta)=\left\{
{\matrix{
{1} & {\rm if} & {0 \le \theta < \pi/2}
\cr
{-1} & {\rm if} & {\pi/2 \le \theta \le \pi.}
} } \right.
\end{equation}
This bound is represented by a thick line in Fig.~\ref{FS02}.

The next problem is how to experimentally achieve this bound.
Given a particular set of local measurements (i.e., a particular
value of $\theta$), which state should we prepare to obtain the
quantum upper and lower bounds? It can be easily seen that the
quantum upper bound is reached by the maximally entangled states
$|\varphi (\xi) \rangle$ given by (\ref{frontierstate}), taking
\begin{equation}
\xi = \frac{1}{2} \left( \theta-g(\theta) \arccos
\left\{\left[1+\sin^2 (2 \theta)\right]^{-1/2}\right\}\right).
\label{xi}
\end{equation}
The quantum lower bound is obtained just by introducing a minus
sign inside the arc cosine in (\ref{xi}). No other quantum state
gives values outside these bounds.

\begin{figure}
\centerline{\includegraphics[width=6.8cm]{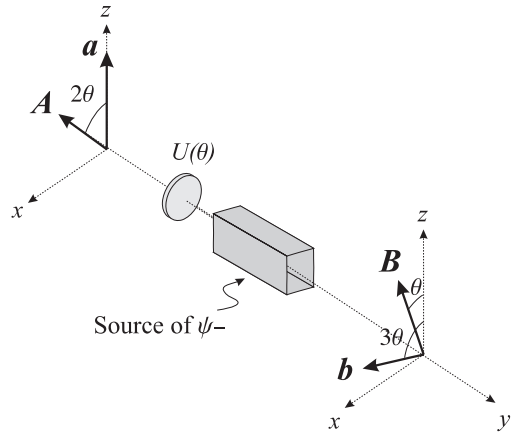}}
\caption{\label{Setup02}
Scheme of the experiment to test the
bounds of quantum correlations. It consists of a source of two-qubit singlet
states $|\psi^-\rangle$, a unitary operation $U(\theta)$ [given by
(\ref{U}) and (\ref{xi})] on qubit $I$, and the local measurements
$A(\theta)$ [given by (\ref{A})] and (alternatively) $a(\theta)$
[given by (\ref{a})] on qubit $I$, and $B(\theta)$ [given by
(\ref{B})] and (alternatively) $b(\theta)$ [given by (\ref{b})] on
qubit $II$. The experiment consists of a
set of $N$ Bell's inequality-type tests for $N$ different values
of $\theta$ in the range $[0, \pi]$.}
\end{figure}

For practical purposes, it is useful to realize that the required
initial states can be prepared using a source of singlet states
\begin{equation}
|\psi^-\rangle = {1 \over \sqrt{2}} \left( |01\rangle-|10\rangle
\right),
\end{equation}
and applying a suitable unitary transformation $U(\theta)$ to
particle $I$. This follows from the fact that, for any $\xi$,
\begin{equation}
|\varphi (\xi) \rangle = U(\xi) \otimes \id \, |\psi^-\rangle,
\end{equation}
where
\begin{equation}
U(\xi) = \left( {\matrix{ {\sin\xi}&{-\cos\xi}\cr
{\cos\xi}&{\sin\xi}\cr }} \right), \label{U}
\end{equation}
and $\id$ is the identity matrix. Therefore, the setup required
to test the quantum bound $F(\theta)$ is illustrated in
Fig.~\ref{Setup02}. It consists of a source of two-qubit singlet
states $|\psi^-\rangle$, a unitary operation $U(\theta)$ [given by
(\ref{U}) and (\ref{xi})] on qubit $I$, and the local measurements
$A(\theta)$ [given by (\ref{A})] and (alternatively) $a(\theta)$
[given by (\ref{a})] on qubit $I$, and $B(\theta)$ [given by
(\ref{B})] and (alternatively) $b(\theta)$ [given by (\ref{b})] on
qubit $II$. A systematical test of the bounds of quantum
correlations can be achieved by performing $N$ Bell's
inequality-type tests, each for a particular value of $\theta$
(i.e., for a particular choice of local observables $A(\theta)$,
$a(\theta)$, $B(\theta)$, and $b(\theta)$, and a particular state
$|\varphi(\theta)\rangle$), covering the range $0 \le \theta \le
\pi$.

%%%%%%%%%%%%%%%%%%%%%%%%%%%% Figure 2 %%%%%%%%%%%%%%%%%%%%%%%%%%%%%

%%%%%%%%%%%%%%%%%%%%%%%%%%%%%%%%%%%%%%%%%%%%%%%%%%%%%%%%%%%%%%%%%%%%

In order to make the CHSH inequality (\ref{CHSH}) useful for real
experiments, it is common practice to translate it into the
language of joint probabilities. This leads to the Clauser-Horne
(CH) inequality~\cite{CH74,Mermin95}:
\begin{eqnarray}
-1 & \le & P\left(A=1,B=1\right) - P\left(A=1,b=-1\right) \nonumber \\
 & & +P\left(a=1,B=1\right) + P\left(a=1,b=-1\right) \nonumber \\
 & & -P\left(a=1\right) - P\left(B=1\right) \le 0.
\label{CH}
\end{eqnarray}
It can be easily seen that the bounds $l$ of the CHSH inequality
(\ref{CHSH}) are transformed into the bounds $(l-2)/4$ of the CH
inequality (\ref{CH}). Therefore, the local-realistic bound in the
CH inequality is 0, and Tsirelson's bound is $(\sqrt{2}-1)/2$.
Analogously, if we calculate the values of the CH operator
\begin{eqnarray}
{\rm CH} & = & P_\rho \left(A=1,B=1\right) - P_\rho \left(A=1,b=-1\right) \nonumber \\
 & & +P_\rho \left(a=1,B=1\right) + P_\rho \left(a=1,b=-1\right) \nonumber \\
 & & -P_\rho \left(a=1\right) - P_\rho \left(B=1\right),
\label{CHoperator}
\end{eqnarray}
for the states $|\varphi(\xi)\rangle$, we obtain a figure which
looks like Fig.~\ref{FS02} but has a different scale in the
vertical axis \{i.e., instead of points with coordinates $(\theta,
{\rm CHSH})$, we obtain points with coordinates $\left[\theta, ({\rm
CHSH}-2)/4\right]$\}.

This systematic test of the bounds of quantum correlations can be
performed with current technology. A physical system particularly
suitable for its implementation consists of pairs of photons
entangled in polarization produced by degenerate type-II
parametric down-conversion~\cite{KMWZSS95,WJSHZ98}. In this
system, the role of spin observables is played by polarization
observables which are particularly adequate due to the
availability of high efficiency polarization-control elements and
the relative insensitivity of most materials to birefringent
thermally induced drifts. An essential advantage of this system is
that it allows Bell's inequality-type tests under strict
space-like separations~\cite{WJSHZ98} (however, current detector
efficiencies do not allow these experiments to elude the so-called
detection loophole~\cite{Grangier01}).

%%%%%%%%%%%%%%%%%%%%%%%% Benefits %%%%%%%%%%%%%%%%%%%%%%%%

Apart from providing a systematical way to experimentally verify
set of extreme nonclassical predictions of quantum mechanics, two
kind of benefits are expected from the proposed experiment.

{\em Identifying the source of experimental errors in tests of
Bell's inequalities.---}Since the Hilbert space structure of
quantum mechanics is not used in the derivation of Bell's
inequalities, the main conclusion of an experimental violation of
a Bell's inequality is clear: The experimental results are
incompatible with local realism. The role of quantum mechanics in
a test of Bell's inequalities is to tell us which physical system
we should prepare and in which directions we should orientate our
polarizers or Stern-Gerlach devices. However, quantum mechanics
does not only tell us this; it also predicts a specific result for
the experiment. The point is that this specific prediction relies
on some additional assumptions. Some of these assumptions are
related to the inefficiencies of our preparations and detectors.
Other assumptions concern the adequacy of the quantum-mechanical
description of the experiment. The failure of each of these two
kinds of assumptions has a different effect on the results. For
instance, if the state we have prepared is a Werner
state~\cite{Werner89} such as $\rho=(1-\epsilon) |\psi^-\rangle
\langle \psi^-|+\epsilon \id/4$ with $0 < \epsilon \ll 1$, instead
of $|\psi^-\rangle$, then the quantum prediction for the proposed
experiment is not~$F(\theta)$ given by (\ref{quantumbound}), but a
curve very close to $F(\theta)$ comprised between the quantum
bounds. In other words, in this case the distance between the
theoretical prediction assuming the $|\psi^-\rangle$ and the
experimental result is not significatively sensitive to the value
of parameter $\theta$. However, if we have assumed that the
measured local observables are accurately described by a
two-dimensional Hilbert space, but that a more adequate
description would require a higher dimensional Hilbert space then,
even if both quantum predictions were similar for some value of
$\theta$ and both are comprised between the quantum bounds, the
distance between them will be very sensitive to the value of
$\theta$.

{\em Searching for correlations beyond those predicted by quantum
mechanics.---}The proposed experiment can be modified to search for
hypothetical correlations beyond those predicted by quantum
mechanics (i.e., superquantum correlations in the extended sense
mentioned above). We do not have any plausible theory which
predicts these correlations and helps us design an experiment
showing violations of the inequalities~(\ref{Masanes}). However,
by the very definition of the bounds $F(\theta)$, for a given set
of local observables, no quantum state can give values outside the
bounds $F(\theta)$. To verify this for any fixed set of
alternative local observables, we can randomly modify the state
emitted by the source. Quantum mechanics predicts that there are
no results outside these bounds. The existence of experimental
results outside these bounds would mean that there are procedures
for preparing physical systems which are not described by any
quantum state and, therefore, that quantum mechanics is incomplete.\\

%%%%%%%%%%%%%%%%%%%%%% Acknowledgements %%%%%%%%%%%%%%%%%%%%%%

I thank M.~Bourennane, L.~Masanes, and H.~Weinfurter for helpful
discussions, I.~Pitowsky and B.~S.~Tsirelson for references, and
the Spanish Ministerio de Ciencia y Tecnolog\'{\i}a Grant
No.~BFM2002-02815 and the Junta de Andaluc\'{\i}a Grant
No.~FQM-239 for support.

%%%%%%%%%%%%%%%%%%%%%%%%% References %%%%%%%%%%%%%%%%%%%%%%%%%

%%%%%%%%%%%%%%%%%%%%%%%%%%%%%%%%%%%%%%%%%%%%%%%%%%%%%%%%%%%%%%%%%%%%


\begin{thebibliography}{99}

%%%%%%%%%%%%%%%%%%%%%%%%%% Loopholes %%%%%%%%%%%%%%%%%%%%%%%%%%

\bibitem{Aspect99}
A. Aspect,
%``Bell's inequality test: More ideal than ever'',
%{\em Nature} {\bf 398}, 6724, 189-190 (1999).
Nature (London) {\bf 398}, 189 (1999).

\bibitem{Grangier01}
P. Grangier,
%``Quantum physics: Count them all'',
%{\em Nature} {\bf 409}, 6822, 774-775 (2001).
Nature (London) {\bf 409}, 774 (2001).

%%%%%%%%%%%%%%%%%%%%%%%%%% EPR, Bell %%%%%%%%%%%%%%%%%%%%%%%%%%

\bibitem{EPR35}
A. Einstein, B. Podolsky, and N. Rosen,
%``Can quantum-mechanical description of physical reality
%be considered complete?'',
Phys. Rev. {\bf 47}, 777 (1935).

\bibitem{Bell64}
J.S. Bell,
%``On the Einstein-Podolsky-Rosen paradox'',
Physics (Long Island City, NY) {\bf 1}, 195 (1964).

%%%%%%%%%%%%%%%%%%%%%%%%% Experiments %%%%%%%%%%%%%%%%%%%%%%%%%

\bibitem{ADR82}
A. Aspect, J. Dalibard, and G. Roger,
%``Experimental test of Bell's inequalities using time-varying
%analyzers'',
Phys. Rev. Lett. {\bf 49}, 1804 (1982).

\bibitem{SA88}
Y.H. Shih and C.O. Alley,
%``New type of Einstein-Podolsky-Rosen-Bohm experiment using pairs
%of light quanta produced by optical parametric down conversion'',
%{\em Phys. Rev. Lett.} {\bf 61}, 26, 2921-2924 (1988).
Phys. Rev. Lett. {\bf 61}, 2921 (1988).

\bibitem{OM88}
Z.Y. Ou and L. Mandel,
%``Violation of Bell's inequality
%and classical probability in a two-photon correlation experiment'',
%{\em Phys. Rev. Lett.} {\bf 61}, 1, 50-53 (1988).
Phys. Rev. Lett. {\bf 61}, 50 (1988).

\bibitem{OPKP92}
Z.Y.~Ou, S.F.~Pereira, H.J.~Kimble, and K.C.~Peng,
%``Realization of the Einstein-Podolsky-Rosen paradox for continuous
%variables'',
%{\em Phys. Rev. Lett.} {\bf 68}, 25, 3663-3666 (1992).
Phys. Rev. Lett. {\bf 68}, 3663 (1992).

\bibitem{TRO94}
P.R.~Tapster, J.G.~Rarity, and P.C.M.~Owens,
%``Violation of Bell's inequality over 4 km of optical fiber'',
%{\em Phys. Rev. Lett.} {\bf 73}, 14, 1923-1926 (1994).
Phys. Rev. Lett. {\bf 73}, 1923 (1994).

\bibitem{KMWZSS95}
P.G.~Kwiat, K.~Mattle, H.~Weinfurter, A.~Zeilinger,
A.V.~Sergienko, and Y.~Shih,
%``New high-intensity source of polarization-entangled photon pairs'',
%{\em Phys. Rev. Lett.} {\bf 75}, 24, 4337-4341 (1995).
Phys. Rev. Lett. {\bf 75}, 4337 (1995).

\bibitem{TBZG98}
W.~Tittel, J.~Brendel, H.~Zbinden, and N.~Gisin,
%``Violation of Bell inequalities by photons more than 10 km apart'',
%{\em Phys. Rev. Lett.} {\bf 81}, 17, 3563-3566 (1998);
Phys. Rev. Lett. {\bf 81}, 3563 (1998).

\bibitem{WJSHZ98}
G.~Weihs, T.~Jennewein, C.~Simon, H.~Weinfurter, and A.~Zeilinger,
%``Violation of Bell's inequality under strict
%Einstein locality conditions'',
Phys. Rev. Lett. {\bf 81}, 5039 (1998).

\bibitem{RKVSIMW01}
M.A.~Rowe, D.~Kielpinski, V.~Meyer, C.A.~Sackett, W.M.~Itano,
C.~Monroe, and D.J.~Wineland,
%``Experimental violation of a Bell's inequality
%with efficient detection'',
Nature (London) {\bf 409}, 791 (2001).

%%%%%%%%%%%%%%%%%%%%%%%%% CHSH %%%%%%%%%%%%%%%%%%%%%%%%%%%%%

\bibitem{Fine82a}
A. Fine,
%A.I. Fine,
%``Hidden variables, joint probability, and the Bell inequalities'',
%{\em Phys. Rev. Lett.} {\bf 48}, 5, 291-295 (1982).
Phys. Rev. Lett. {\bf 48}, 291 (1982).

\bibitem{GM82}
A. Garg and N.D. Mermin,
%``Comment on
%`Hidden variables, joint probability, and the Bell inequalities'\,'',
%{\em Phys. Rev. Lett.} {\bf 49}, 3, 242 (1982).
Phys. Rev. Lett. {\bf 49}, 242 (1982).

\bibitem{Fine82b}
A. Fine,
%A.I. Fine,
%``Fine responds'',
%{\em Phys. Rev. Lett.} {\bf 49}, 3, 243 (1982).
Phys. Rev. Lett. {\bf 49}, 243 (1982).

\bibitem{WW01}
R.F. Werner and M.M. Wolf,
%``All-multipartite Bell-correlation inequalities
%for two dichotomic observables per site'',
%{\em Phys. Rev. A} {\bf 64}, 3, 032112 (2001).
Phys. Rev. A {\bf 64}, 032112 (2001).

\bibitem{CHSH69}
J.F.~Clauser, M.A.~Horne, A.~Shimony, and R.A.~Holt,
%``Proposed experiment to test local hidden-variable theories'',
Phys. Rev. Lett. {\bf 23}, 880 (1969).

%%%%%%%%%%%%%%%%%%%%%%%% Tsirelson %%%%%%%%%%%%%%%%%%%%%%%%%

\bibitem{Tsirelson80}
B.S. Tsirelson,
%``Quantum generalizations of Bell's inequality'',
Lett. Math. Phys. {\bf 4}, 93 (1980).

\bibitem{Landau87}
L.J. Landau,
%``On the violation of Bell's inequality in quantum theory'',
Phys. Lett. A {\bf 120}, 54 (1987).

\bibitem{BMR92}
S.L. Braunstein, A. Mann, and M. Revzen,
%``Maximal violation of Bell inequalities for mixed states'',
Phys. Rev. Lett. {\bf 68}, 3259 (1992).

%%%%%%%%%%%%%%%% Superquantum correlations %%%%%%%%%%%%%%%%

\bibitem{PR94}
S. Popescu and D. Rohrlich,
%``Quantum nonlocality as an axiom'',
Found. Phys. {\bf 24}, 379 (1994).

%%%%%%%%%%%%%%%%%% Uffink's inequalities %%%%%%%%%%%%%%%%%%

\bibitem{Uffink02}
J. Uffink,
%``Quadratic Bell inequalities as tests for multipartite entanglement'',
%{\em Phys. Rev. Lett.} {\bf 88}, 23, 230406 (2002).
Phys. Rev. Lett. {\bf 88}, 230406 (2002).

%%%%%%%%%%%% Convex set of quantum correlations %%%%%%%%%%%%

\bibitem{Landau88}
L.J. Landau,
%``?'',
%{\em Found. Phys.} {\bf 18}, ?, 449-? (1988).
Found. Phys. {\bf 18}, 449 (1988).

\bibitem{Tsirelson93}
B.S. Tsirelson,
%``Some results and problems on quantum Bell-type inequalities'',
%{\em Hadronic J. Supplement} {\bf 8}, 4, 329-345 (1993).
Hadronic J. Supplement {\bf 8}, 329 (1993).

\bibitem{Masanes03}
L. Masanes,
%``Necessary and sufficient condition for quantum-generated correlations'',
quant-ph/0309137.

\bibitem{FS03}
S. Filipp and K. Svozil,
%``Testing the bounds on quantum probabilities'',
quant-ph/0306092 [Phys. Rev. Lett. (to be published)].

%%%%%%%%%%%%% Classical correlation polytopes %%%%%%%%%%%%%

\bibitem{Pitowsky86}
I. Pitowsky,
J. Math. Phys. {\bf 27}, 1556 (1986).

\bibitem{Pitowsky89}
I. Pitowsky,
{\em Quantum Probability-Quantum Logic}
(Springer, Berlin, 1989).

%%%%%%%%%%%%%%%%%%%%%%%%%%% CH %%%%%%%%%%%%%%%%%%%%%%%%%%%

\bibitem{CH74}
J.F. Clauser and M.A. Horne,
%``Experimental consecuences of objective local theories'',
Phys. Rev. D {\bf 10}, 526 (1974).

\bibitem{Mermin95}
N.D. Mermin,
%``The best version of Bell's theorem'',
%in {\em Fundamental Problems in Quantum Theory:
%A Conference Held in
%Honor of Professor John A. Wheeler},
%edited by D.M. Greenberger and A. Zeilinger,
Ann. N. Y. Acad. Sci. {\bf 755}, 616 (1995).

%%%%%%%%%%%%%%%%%%%%%% Werner states %%%%%%%%%%%%%%%%%%%%%%

\bibitem{Werner89}
R.F. Werner,
%``Quantum states with Einstein-Podolsky-Rosen correlations
%admitting a hidden-variable model'',
%{\em Phys. Rev. A} {\bf 40}, 8, 4277-4281 (1989).
Phys. Rev. A {\bf 40}, 4277 (1989).

%%%%%%%%%%%%%%%%%%%%%%%%%%%%%%%%%%%%%%%%%%%%%%%%%%%%%%%%%%%%%%%%%%%%

\end{thebibliography}
\end{document}